\documentclass[lettersize,journal]{IEEEtran}
\usepackage{amsmath,amsfonts}
\usepackage{algorithmic}
\usepackage{algorithm}
\usepackage{array}
\usepackage[caption=false,font=normalsize,labelfont=sf,textfont=sf]{subfig}
\usepackage{textcomp}
\usepackage{xcolor}
\usepackage{algorithmic}
\usepackage{algorithm}
\usepackage{stfloats}
\usepackage{url}
\usepackage[hidelinks]{hyperref}
\usepackage{booktabs}
\usepackage{tikz}
\usepackage{capt-of}
\usepackage{svg}
\usetikzlibrary{arrows.meta,positioning,fit,calc}
\usepackage{multirow}
\usepackage{cleveref} 
\usepackage{bm}
\usepackage{verbatim}
\usepackage{graphicx}
\usepackage{cite}
\usepackage{array}
\newcolumntype{M}{>{\raggedright\arraybackslash}p{3.25cm}}
\begin{document}

\title{
Positive-Incentive Noise Predictor for \\ Adversarial Purification in Speaker Verification
}
\author{Yibo Bai,~\IEEEmembership{}
        Sizhou Chen,~\IEEEmembership{}
        Michele Panariello,~\IEEEmembership{}
        Hao Ma,~\IEEEmembership{}
        Xiao-Lei Zhang,~\IEEEmembership{}\\
        Xuelong Li,~\IEEEmembership{}
        Massimiliano Todisco,~\IEEEmembership{}
        Nicholas Evans~\IEEEmembership{}

\thanks{Yibo Bai, Michele Panariello, Massimiliano Todisco and Nicholas Evans are with the Audio Security and Privacy Group, EURECOM, 06904 Biot, France (e-mail: yibo.bai@eurecom.fr, michele.panariello@eurecom.fr, massimiliano.todisco@eurecom.fr, nicholas.evans@eurecom.fr).}
\thanks{Sizhou Chen is with the School of Computer Science, The University of Sydney, Sydney, NSW 2006, Australia (e-mail: szchen1005@gmail.com).}
\thanks{Hao Ma and Xiao-Lei Zhang are with the School of Marine Science and Technology, Northwestern Polytechnical University, Xi’an 710072, China, also with the Institute of Artificial Intelligence (TeleAI), China Telecom, P. R. China, and also with the Research and Development Institute of Northwestern Polytechnical University in Shenzhen, China (e-mail: haoma@mail.nwpu.edu.cn, xiaolei.zhang@nwpu.edu.cn).}
\thanks{Xuelong Li is with the Institute of Artificial Intelligence
(TeleAI), China Telecom, P. R. China (e-mail: xuelong\_li@ieee.org).}
}

\markboth{Journal of \LaTeX\ Class Files,~Vol.~14, No.~8, August~2021}%
{Shell \MakeLowercase{\textit{et al.}}: A Sample Article Using IEEEtran.cls for IEEE Journals}


\maketitle

\begin{abstract}

Modern automatic speaker verification (ASV) systems are vulnerable to adversarial perturbations. Diffusion-based purification has recently shown strong effectiveness against such perturbations, but its reverse denoising process requires iterative sampling and leads to high inference latency. We find that the forward noising process provides most of the robustness gain. Motivated by this observation, we reformulate adversarial purification as a learnable noising problem, and propose the Positive-Incentive Noise Predictor (PnP), the first framework that explicitly introduces positive-incentive noise ($\pi$-noise) into the purification task. PnP learns input-adaptive $\pi$-noise and mixes it with the input to improve the robustness of downstream ASV systems. Experiments on four advanced ASV backbones show that PnP effectively defends against adversarial attacks while preserving performance on natural speech. Compared with representative purification baselines, the proposed framework provides a competitive balance among defense effectiveness, impact on genuine utterances, and inference efficiency under white-box, black-box, and defender-aware adaptive attacks, with a real-time factor as low as 0.014. Moreover, PnP can be cascaded with a diffusion denoiser to further improve the perceptual quality of purified utterances. Code and purified audio examples are available at \url{https://eurecom-asp.github.io/pnp/}

\end{abstract}


\section{Introduction}

\IEEEPARstart{A}{utomatic} speaker verification (ASV) aims to authenticate users by comparing speaker representations extracted from speech signals \cite{bai2021speaker}. As for any biometric errors may cause genuine users to be rejected or impostors to be accepted. Recent deep speaker embeddings and large-scale pretrained speech encoders have substantially advanced ASV performance, including wav2vec 2.0 \cite{baevski2020wav2vec}, WavLM \cite{chen2022wavlm}, and Whisper \cite{radford2023robust}. Despite these advances, state-of-the-art ASV systems remain vulnerable to spoofing and adversarial attacks. Spoofing attacks use synthetic or converted speech to impersonate a target speaker, motivating spoofing countermeasures such as those developed within the context of ASVspoof challenges \cite{wang2024asvspoof}. Adversarial attacks instead add small, human-imperceptible perturbations to speech signals to manipulate downstream decisions \cite{szegedy2014intriguing}.

Adversarial attacks have been shown to be effective against ASV systems while maintaining high perceptual quality, often with signal-to-noise ratios (SNRs) above 30 dB \cite{villalba2020x}. Such adversarial perturbations can induce the false rejection of genuine speakers or the false acceptance of impostors. By optimizing against the verification objective, adversarial attack perturbations are crafted to mislead the target model. Most attacks are gradient-based \cite{goodfellow2014explaining,kurakin2018adversarial,madry2018towards} and typically assume white-box access to the victim model. In gray-box or black-box settings, such access to model architecture and parameters is limited or unavailable. In these settings, transfer-based attack perturbations are crafted using surrogate models \cite{li2020adversarial,yao2024interpretable} and query-based attack perturbations are optimized with gradient estimation \cite{chen2021real}. Adversarial threats have also been extended to more realistic settings, including streaming input \cite{li2020advpulse}, over-the-air scenarios \cite{wang2026over}, and text-dependent ASV \cite{sankala2025adversarial}.

Recent defense methods against adversarial attacks have explored proactive and passive defenses \cite{wu2023defender}. Proactive defense modifies the model to improve robustness before deployment. For example, adversarial training incorporates adversarial examples during training \cite{lan2022adversarial}, which requires generating adversarial examples and retraining for each target ASV model. Passive defense includes adversarial example detection and adversarial purification. The former aims to distinguish genuine inputs from adversarial utterances. For example, learnable mask detector (LMD) \cite{chen2023lmd} masks selected time-frequency regions and compares embedding similarity differences before and after masking to detect adversarial examples. Self-supervised models have also been applied to both detection and purification \cite{wu2021improving}. Mask diffusion detector (MDD) \cite{bai2025mdd} masks the Mel-spectrogram and uses a diffusion model to reconstruct masked regions. However, in detection-based methods, such reconstructed outputs are often too distorted for subsequent ASV scoring, and the detection threshold may vary across models or input distributions.

This paper focuses on adversarial purification, which passively processes every input to suppress potential adversarial perturbations and maintain performance on genuine utterances. Early purification approaches used simple signal processing, such as filtering \cite{wu2020defense} or additive noise \cite{chang2021defending}, but may easily degrade perceptual quality. More recently, neural audio codecs have been explored as purifiers, as their compression and decompression processes can remove adversarial perturbations \cite{chen2024neural}. Generative approaches have also shown strong performance. Diffusion-based adversarial purification (DAP) \cite{bai2024diffusion} was the first to introduce diffusion models for ASV purification, and textual-driven adversarial purification (TDAP) \cite{chen2024textual} further adds text conditioning into the purification process. Despite their strong defense performance, diffusion-based purification methods typically require step-by-step sampling, which leads to high latency. Two-stage diffusion-based purification (TDP) \cite{bai2024adversarial} improves efficiency by purifying in a Mel-spectrogram latent space, but still relies on a vocoder to reconstruct waveforms, which degrades audio quality. Compared with proactive and detection-based passive defenses,  purification is more attractive in few-shot or zero-shot settings since it does not require model modification or threshold calibration.

A diffusion-based purifier consists of two stages. The forward noising process gradually corrupts the input toward pure Gaussian noise, and the reverse denoising process iteratively predicts and removes the injected noise. From this viewpoint, the forward noising process alone is conceptually close to additive-noise defense methods \cite{chang2021defending}, which also perturb the input with noise of fixed standard deviation. This observation naturally raises a question: is the reverse denoising process necessary for effective adversarial purification?

\begin{figure*}[t]
    \centering
    \includegraphics[width=0.9\textwidth]{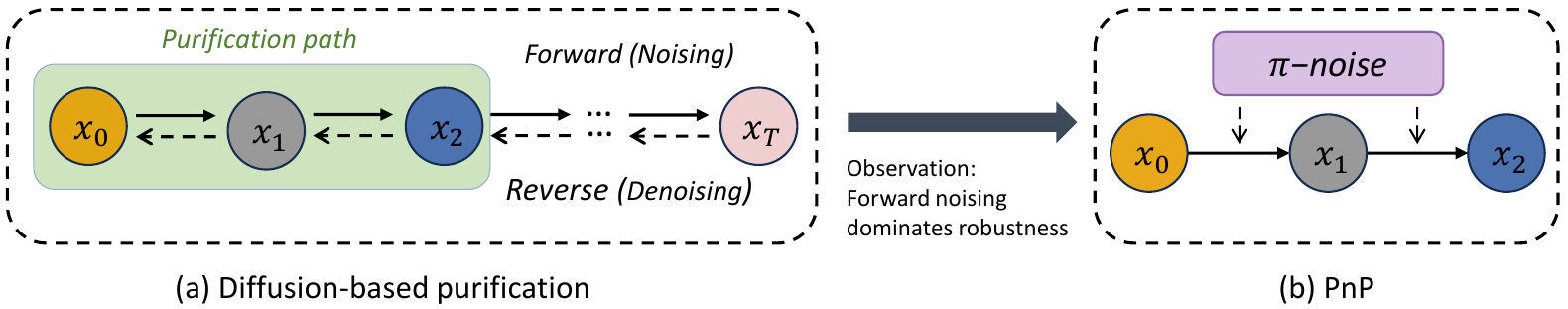}
    \caption{From diffusion-based purification to PnP. 
Diffusion models for generative tasks perform a full forward process from $x_0$ to $x_T$, followed by reverse sampling back to $x_0$. 
In contrast, diffusion-based purification only relies on a partial forward and reverse process (an example of $x_0 \rightarrow x_2 \rightarrow x_0$ is highlighted in green). 
Motivated by the observation that most robustness gains actually come from the forward process, PnP reformulates purification as a learned forward noising procedure with task-beneficial positive-incentive noise ($\pi$-noise). The right part shows the diffusion-style PnP variant.}
    \label{fig:intro-puri}
\end{figure*}

If most robustness gains already come from forward noising, adversarial purification for ASV can instead be reformulated as a learned forward noising problem, rather than relying on a full diffusion-based purification pipeline. Our preliminary experiments support this hypothesis. Motivated by this observation, we propose a unified input-adaptive Positive-Incentive Noise Predictor (PnP) framework, which learns task-beneficial noise as a front-end purification module for speaker verification. Fig.~\ref{fig:intro-puri} illustrates this shift in perspective. Instead of relying on reverse denoising, we focus on the forward process and investigate whether the injected noise can be learned to be task-beneficial. Under this framework, \textbf{PnP-Gaussian} provides a simple additive formulation, while \textbf{PnP-Diff} provides a diffusion-style formulation aligned with the forward noising process of a diffusion model. Moreover, the diffusion-style variant can be optionally cascaded with a diffusion denoiser for audio quality enhancement.

The main contributions are summarized below.
\begin{itemize}
    \item We revisit diffusion-based audio adversarial purification and show that robustness gains come largely from the forward noising process rather than the more expensive reverse denoising process. This provides a new perspective for simplifying diffusion-based defenses in ASV.

    \item Based on this observation, we propose the first framework that uses positive-incentive noise ($\pi$-noise) in the adversarial purification task. Our PnP method learns input-adaptive $\pi$-noise as a front-end purifier for ASV. Within this unified framework, PnP-Gaussian serves as an adaptive version of additive noising, while PnP-Diff replaces the fixed Gaussian forward noise in diffusion purification with learned $\pi$-noise.

\item Extensive experiments show that the proposed framework consistently improves adversarial robustness in ASV under white-box, black-box, and defender-aware adaptive attacks. The 1-step and 2-step PnP-Diff variants achieve state-of-the-art robustness across different attack settings with a real-time factor (RTF) of 0.014. In addition, cascading PnP-Diff with a diffusion denoiser further improves perceptual quality, reaching a WB-PESQ of 3.591 and an SI-SDR of 21.14 dB on adversarial VoxCeleb data.
\end{itemize}

\section{Background}

\subsection{Automatic Speaker Verification}
An ASV system aims to verify the speaker identity of an input test utterance $x^{\mathrm{test}}$. Given a pair of waveforms $(x^{\mathrm{test}}, x^{\mathrm{enroll}})$, where $x^{\mathrm{enroll}}$ is an enrollment utterance from a known speaker, the ASV system extracts speaker embeddings and computes a similarity score $s(x^{\mathrm{test}}, x^{\mathrm{enroll}})$ between them. The verification decision is made by comparing the similarity score with a threshold.

Common ASV embedding extractors include x-vector~\cite{snyder2018x}, ECAPA-TDNN~\cite{desplanques2020ecapa}, and ResNet-based encoders~\cite{zeinali2019but}. These systems typically take logarithmic filterbanks (LogFBank) as input. Recently, self-supervised learning (SSL) speech models such as wav2vec~2.0~\cite{baevski2020wav2vec} and WavLM~\cite{chen2022wavlm} have also been adopted as upstream feature extractors, replacing conventional LogFBank features~\cite{chen2022large}.

\subsection{Adversarial attacks in ASV}
Adversarial attacks were first introduced in the image classification domain~\cite{szegedy2014intriguing}. Despite appearing visually similar to original genuine examples, adversarial images can mislead well-trained classifiers. Such examples are typically crafted to cross the classification boundary while being constrained by an $\ell_p$ norm to remain imperceptible. Subsequent works proposed stronger optimization-based attacks and robustness evaluations, such as FGSM~\cite{goodfellow2014explaining}, PGD~\cite{madry2018towards}, and CW~\cite{carlini2017towards}. These attacks can be conducted in a white-box setting, where the attacker has full model knowledge, or in a black-box setting, where the attacker has no access to the internal parameters of the target model and may then rely on either transfer-based or query-based methods.

In the ASV setting, the attacker aims to craft a perturbed adversarial test utterance
$x^{\mathrm{adv}} = x^{\mathrm{test}} + \delta$ with extremely small $\|\delta\|_p$.
Although $x^{\mathrm{adv}}$ still sounds identical to $x^{\mathrm{test}}$, it may cause false accepts for unauthorized speakers or false rejects for genuine speakers.

In this paper, we consider MI-FGSM~\cite{dong2018boosting} and PGD~\cite{madry2018towards} as two strong,  gradient-based attacks.

\subsubsection{MI-FGSM}
Momentum iterative fast gradient sign method (MI-FGSM) is an iterative $\ell_\infty$ attack with momentum.
Let $\mathcal{J}(x)$ denote the attack objective. Starting from $x_0=x^{\mathrm{test}}$ and $g_0=\mathbf{0}$, MI-FGSM iterates
\begin{align}
g_{k+1} &= \mu\, g_k + \frac{\nabla_x \mathcal{J}(x_k)}{\|\nabla_x \mathcal{J}(x_k)\|_1}, \\
x_{k+1} &= \Pi_{\mathcal{B}_\infty}\!\left(x_k + \alpha\,\cdot \mathrm{sign}(g_{k+1})\right),
\end{align}
where $\alpha$ is the step size, $\mu$ is the momentum decay, and $\Pi_{\mathcal{B}_p}(\cdot)$ denotes the projection onto a norm sphere $\ell_p$ that constrains the perturbation magnitude.

\subsubsection{PGD}
Projected gradient descent (PGD) is a multi-step first-order attack with projection.
Let $\mathcal{J}(x)$ denote the attack objective. In our experiments, we evaluate PGD-$\ell_\infty$ and PGD-$\ell_2$ attacks under different types of norm constraints. Starting from $x_0=x^{\mathrm{test}}$, PGD iterates
\begin{align}
\text{PGD-}\ell_\infty:\quad
x_{k+1} &= \Pi_{\mathcal{B}_\infty}\!\left(x_k+\alpha\,\cdot \mathrm{sign}\!\left(\nabla_x \mathcal{J}(x_k)\right)\right),\\
\text{PGD-}\ell_2:\quad
x_{k+1} &= \Pi_{\mathcal{B}_2}\!\left(x_k+\alpha\, \cdot \frac{\nabla_x \mathcal{J}(x_k)}{\|\nabla_x \mathcal{J}(x_k)\|_2}\right).
\end{align}

\subsection{Diffusion-based purification}
To protect models from adversarial attacks, purification methods aim to remove adversarial perturbations from the input. DiffPure~\cite{nie2022diffusion} was first proposed to defend image classifiers with pretrained diffusion models. As a diffusion model consists of a forward noising process and a reverse denoising process, given an input sample $x_0$, the forward process can be formulated as
\begin{equation}
\label{eq:ddpm_forawrd}
x_t = \sqrt{\bar{\alpha}_t}\,x_0 + \sqrt{1-\bar{\alpha}_t}\,\bm{\epsilon},
\end{equation}
where $\bm{\epsilon} \sim \mathcal{N}(0,I)$ is a Gaussian noise and $\bar{\alpha}_t$ at step $t$ is computed by a predefined noise schedule. In the reverse process, the diffusion model is trained to predict an estimated noise $\bm{\epsilon}_\theta$ to remove this noise step by step. Given an adversarial input $x^\mathrm{adv}$ and a chosen purification step $t^\ast$, DiffPure applies the forward process to obtain $x^\mathrm{adv}_{t^\ast}$ before the reverse process.  AudioPure~\cite{wu2023defending} subsequently applies this idea to adversarial audio for speech emotion classification. In the ASV domain, unlike DiffPure and AudioPure, DAP~\cite{bai2024diffusion} is trained on adversarial data and runs the full reverse process directly on $x^\mathrm{adv}$ rather than $x^\mathrm{adv}_{t^\ast}$, which yields better audio quality for purified utterances but requires extra inference time; additionally, DAP requires large-scale adversarial audio. Sample-specific noise injection (SSNI)~\cite{sun2025sample} further replaces the shared purification level $t^\ast$ with a sample-specific $t(x)$ estimated from the score magnitude of a pretrained diffusion model. SSNI is similar to our work in that it also questions the use of a fixed forward noising process. However, SSNI still assumes Gaussian noise in the forward process and retains the reverse denoising procedure, whereas our PnP framework learns task-beneficial noise and makes the denoiser optional.

\section{Positive-Incentive Noise Predictor for
ASV Adversarial Purification}
\label{sec:pnp}

This section first motivates the Positive-Incentive Noise Predictor (PnP) through a preliminary experiment in Section~\ref{subsec:pnp-motivation}. Section~\ref{PnP} then introduces the unified PnP framework and its learning objective. Section~\ref{PnP-Diff+DM} presents its applications to adversarial purification.

\subsection{Motivation from Diffusion-Based Purification}
\label{subsec:pnp-motivation}
\begin{figure}[t]
    \centering
    \includegraphics[width=0.85\columnwidth]{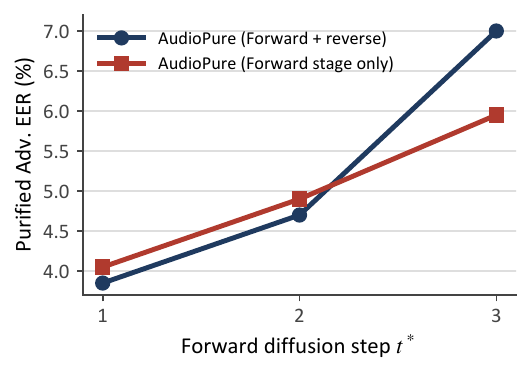}
    \caption{Method motivation from diffusion purification analysis. In this preliminary experiment, we compare the purified adversarial EERs of two AudioPure-based purifiers under 50-step MI-FGSM attack, including the full AudioPure purifier with forward noising plus reverse denoising and its forward-stage-only purifier variant. The small gap between the two curves indicates that forward noising already provides most of the robustness.}
    \label{fig:method-motivation-forward}
\end{figure}

Adversarial purification protects a downstream ASV model from the influence of adversarial perturbations. By revisiting the diffusion-based purification pipeline through AudioPure~\cite{wu2023defending}, we investigate whether most robustness gains already arise from the forward noising stage. In particular, we compare the defense performance of the full AudioPure pipeline with its forward-process-only variant on VoxCeleb1 data attacked by MI-FGSM. As shown in Fig.~\ref{fig:method-motivation-forward}, the forward-only variant achieves purified adversarial EERs that are very close to those of the full pipeline across different forward diffusion steps. More detailed stepwise analysis of AudioPure is provided in Section~\ref{subsec:audiopure}. This motivates us to learn a task-beneficial forward noising process, which we formulate next through a positive-incentive learning objective.

\begin{figure*}[t]
    \centering
    \includegraphics[width=0.95\linewidth]{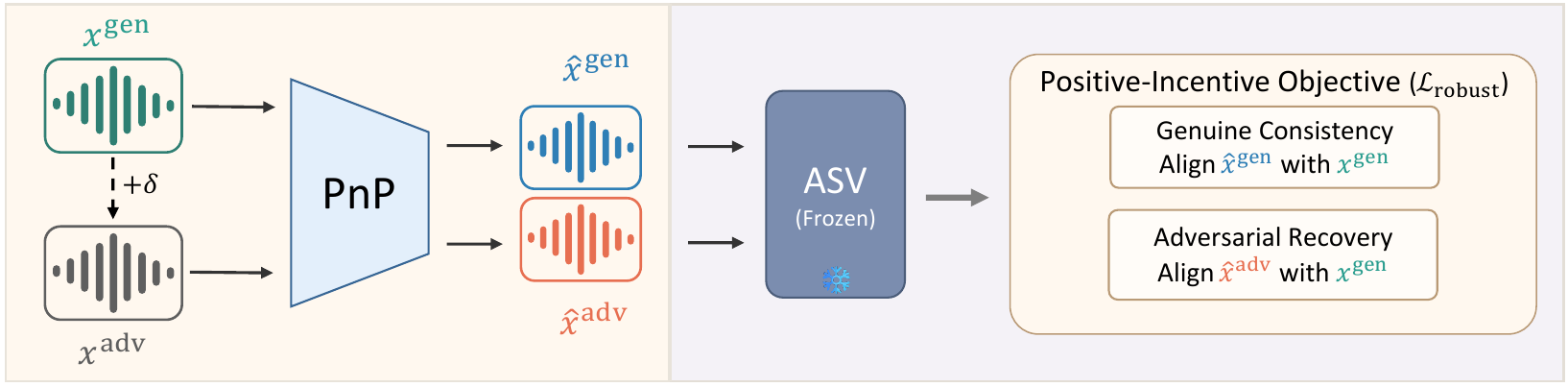}
    \caption{Training pipeline of the proposed positive-incentive noise predictor (PnP). For each input genuine utterance $x^{\mathrm{gen}}$, we generate its adversarial example $x^{\mathrm{adv}}=x^{\mathrm{gen}}+\delta$. PnP processes the genuine and adversarial inputs, and produces purified outputs $\hat{x}^{\mathrm{gen}}$ and $\hat{x}^{\mathrm{adv}}$, which are then fed into a frozen ASV system. The PnP parameters are optimized to increase speaker similarity between the genuine reference and the purified outputs from both branches, thereby preserving performance on unattacked inputs while suppressing adversarial perturbations.}
    \label{fig:train}
\end{figure*}

\subsection{Unified PnP Framework}
\label{PnP}

We now introduce the unified waveform-level framework for ASV adversarial purification and its learning objective. Previous additive-noise defenses~\cite{chang2021defending} and diffusion-based purification methods~\cite{wu2023defending} commonly rely on task-agnostic random noising. In contrast, PnP replaces this random noising with a learnable positive-incentive noising process.

\subsubsection{Framework}
\label{subsubsec:pnp-overview}

Fig.~\ref{fig:train} illustrates the implementation of the ASV-guided PnP framework. Given an input waveform $x$, PnP predicts input-adaptive positive-incentive noise ($\pi$-noise) and combines it with the input to construct a purified waveform $\hat{x}$ for downstream ASV scoring. Unlike ordinary random noise, $\pi$-noise benefits a target task, as it is expected to reduce the uncertainty of the downstream task rather than simply corrupt the input~\cite{li2022positive}. In our ASV purification setting, this means that the learned noise should encourage the purified waveform to preserve speaker information while suppressing adversarial perturbations.

Specifically, we construct a learnable noise predictor $\bm{\varepsilon}_{\bm{\omega}}$ by combining a normalized directional component predicted by a 1-D U-Net $\tilde{\bm{\varepsilon}}$ with Gaussian noise:
\begin{equation}
\label{eq:pnp-mix}
\bm{\varepsilon}_{\bm{\omega}}(x)
= \lambda\,
\frac{\tilde{\bm{\varepsilon}}(x)}{\|\tilde{\bm{\varepsilon}}(x)\|_2}
\;+\;
\sqrt{1-\lambda^2}\;\bm{\epsilon},
\end{equation}
where $\bm{\epsilon}\sim\mathcal{N}(0,\mathbf I)$ and $\lambda\in[0,1]$ controls the mixture ratio.

To incorporate the predicted $\pi$-noise into the input waveform for purification, we adopt a unified weighted mixing formulation:
\begin{equation}
\label{eq:x_pnp}
\hat{x} = w_x\, x + w_n \, \bm{\varepsilon}_{\bm{\omega}}(x),
\end{equation}
where $\hat{x}$ denotes the purified waveform, and $w_x$ and $w_n$ control the contributions of the original signal $x$ and the predicted noise $\bm{\varepsilon}_{\bm{\omega}}(x)$, respectively. Different choices of $w_x$ and $w_n$ lead to different PnP variants, as introduced in Section~\ref{PnP-Diff+DM}. The complete procedures are summarized in three algorithms. Algorithm~\ref{alg:pnp-train} trains the $\pi$-noise predictor guided by a frozen ASV system. Algorithm~\ref{alg:pnp-infer} deploys the trained predictor as a lightweight pre-processing module. Algorithm~\ref{alg:pnp-diffusion} further describes how to train a diffusion denoiser with PnP-generated $\pi$-noise for the optional cascade.

\subsubsection{Positive-Incentive Learning Objective}
\label{AP}

We next formalize the positive-incentive learning objective used to train PnP. In the original formulation, $\pi$-noise is defined as a noise variable that reduces the uncertainty of the downstream task. For a task $\mathcal{T}$, this means that after introducing a $\pi$-noise variable $\bm{\varepsilon}$, the task becomes more predictable, i.e., $H(\mathcal{T}\mid \bm{\varepsilon}) < H(\mathcal{T})$, where $H(\cdot)$ denotes information entropy. If $\bm{\varepsilon}$ does not change the task uncertainty, i.e., $H(\mathcal{T}\mid \bm{\varepsilon}) = H(\mathcal{T})$, then it can be regarded as pure noise. The variational positive-incentive noise (VPN) formulation~\cite{zhang2025variational} further shows that this positive-incentive objective can be optimized through a variational lower bound.

In ASV adversarial purification, this perspective is applied through two directions. The learned noise in PnP should guide both genuine and adversarial inputs in a way that benefits downstream ASV decisions better than random noise. For genuine inputs, PnP should preserve the verification performance of a well-trained ASV system. For adversarial inputs, PnP should suppress adversarial perturbations so that the purified adversarial input moves back toward the genuine reference in the speaker embedding space.

We instantiate the positive-incentive formulation in the ASV setting as follows. Let $\mathcal{T}$ denote the speaker verification task, $x$ denote an input utterance, and $y\in[-1,1]$ denote the ASV verification score. We represent the task by the continuous score $y$ with density $p(y\mid x)$, and define the task entropy as
\begin{equation}
H(\mathcal{T}\mid x)
=\mathbb{E}_{x}
\!\Big[\, - \!\!\int p(y\mid x)\,\log p(y\mid x)\, \mathrm{d}y \Big].
\end{equation}
Let $\bm{\varepsilon}$ be a noise variable sampled from the distribution $\mathcal{D}_{\bm\omega}(x,y)$ parameterized by $\bm\omega$. By definition, the mutual information between $\mathcal{T}$ and $\bm{\varepsilon}$ is
\begin{align}
I(\mathcal{T},\bm{\varepsilon}\mid x)
&=
\mathbb{E}_{x}\;
\mathbb{E}_{\bm{\varepsilon}}\;
\Big[\int p(y,\bm{\varepsilon}\mid x)\,\log \frac{p(y,\bm{\varepsilon}\mid x)}{p(y\mid x)\,p(\bm{\varepsilon}\mid x)}\,\mathrm{d}y \Big].
\label{eq:mi-kl}
\end{align}

Therefore, if $I(\mathcal{T},\bm{\varepsilon}\mid x)>0$, the noise variable $\bm{\varepsilon}$ becomes positive-incentive noise for task $\mathcal{T}$. Following the VPN formulation~\cite{zhang2025variational}, we optimize a variational lower bound of $I(\mathcal{T},\bm{\varepsilon}\mid x)$. For ASV adversarial purification, we consider paired genuine and adversarial inputs in each mini-batch, denoted by $\{(x_i^{\mathrm{gen}},x_i^{\mathrm{adv}})\}_{i=1}^{B}$, and expand the variational lower bound over the two branches as
\begin{equation}
\label{eq:pi-lower-bound}
\begin{aligned}
\mathcal{L}_{\pi}
&\approx
\frac{1}{B}\sum_{i=1}^{B}
\mathbb{E}_{\bm{\varepsilon}}
\big[\,\log q(y_i \mid x_i^{\mathrm{gen}}, \bm{\varepsilon})\,\big] \\
&\quad+
\frac{1}{B}\sum_{i=1}^{B}
\mathbb{E}_{\bm{\varepsilon}}
\big[\,\log q(y_i \mid x_i^{\mathrm{adv}}, \bm{\varepsilon})\,\big],
\end{aligned}
\end{equation}
where $q(y \mid x,\bm{\varepsilon})$ is a variational approximation to $p(y \mid x,\bm{\varepsilon})$, and we assume that $\mathcal{D}_{\omega}(x,y)$ is Gaussian. Since $x_i^{\mathrm{adv}}$ is derived from $x_i^{\mathrm{gen}}$, in practice we parameterize the noise generator using the genuine and adversarial inputs, i.e., $\mathcal{D}_{\omega}(x^{\mathrm{gen}},x^{\mathrm{adv}})$, rather than explicitly relying on the value of $y$.

The two terms in $\mathcal{L}_{\pi}$ correspond to the two desired effects of PnP, preserving ASV performance for genuine inputs and suppressing adversarial perturbations for adversarial inputs. Ideally, the $\pi$-noise predictor $\bm{\varepsilon}_{\bm{\omega}}$ should maximize these two variational terms. However, directly estimating the variational density $q(y \mid x,\bm{\varepsilon})$ is difficult in ASV, since the output is a verification score rather than an explicit probabilistic distribution. We therefore use ASV scores as a task-driven surrogate for the positive-incentive objective. Specifically, we maximize the similarity between the genuine reference and both purified genuine and adversarial outputs. A hinge margin prevents unnecessary optimization once the similarity score is sufficiently high. Other surrogate formulations could also be used to optimize the same objective.

As shown in Fig.~\ref{fig:train}, during PnP training, we first generate $x^{\mathrm{adv}}$ for a given $x^{\mathrm{gen}}$, and then compute $\hat{x}^{\mathrm{gen}}$ and $\hat{x}^{\mathrm{adv}}$ with Eq.~\eqref{eq:x_pnp}. We enforce high speaker similarity between the clean reference $x^{\mathrm{gen}}$ and the two purified outputs, $\hat{x}^{\mathrm{gen}}$ and $\hat{x}^{\mathrm{adv}}$, by minimizing the high-margin hinge loss:
\begin{equation}
\label{eq:robust-loss}
\mathcal{L}_{\mathrm{robust}}
=
\phi_m\!\big(s(x^{\mathrm{gen}},\hat{x}^{\mathrm{gen}})\big)
+
\phi_m\!\big(s(x^{\mathrm{gen}},\hat{x}^{\mathrm{adv}})\big),
\end{equation}
where $\phi_m(s)=\max(0,\,m-s)$ with $m$ chosen close to $1$.

To regularize the magnitude of the learned $\pi$-noise, we penalize the energy:
\begin{equation}
\label{eq:reg-loss}
\mathcal{L}_{\mathrm{reg}}
=
\left\|\bm{\varepsilon}_{\bm{\omega}}(x^{\mathrm{gen}})\right\|_2^2
+
\left\|\bm{\varepsilon}_{\bm{\omega}}(x^{\mathrm{adv}})\right\|_2^2.
\end{equation}

The overall training loss is then given by:
\begin{equation}
\label{eq:total-loss}
\mathcal{L}(\bm{\omega}) = \mathcal{L}_{\mathrm{robust}} + \gamma \mathcal{L}_{\mathrm{reg}},
\end{equation}
where $\gamma>0$ is a weighting coefficient. Algorithm~\ref{alg:pnp-train} summarizes the training procedure.

During the test stage, we deploy the trained PnP as a pre-processing purification module in front of an ASV system. Given a test utterance $x$, PnP predicts task-beneficial noise and mixes it with the input according to Eq.~\eqref{eq:pnp-mix} and Eq.~\eqref{eq:x_pnp} to suppress potential adversarial perturbations. Algorithm~\ref{alg:pnp-infer} summarizes the inference procedure.

\begin{algorithm}[t]
  \caption{Train a positive-incentive noise predictor with a frozen ASV system}
  \label{alg:pnp-train}
  \begin{algorithmic}[1]
      \REQUIRE Training set $\{( x^{\text{gen}}_i, x^{\text{adv}}_i)\}_{i=1}^n$, batch size $B$, and a frozen ASV system.
      \ENSURE Trained $\pi$-noise predictor $\bm{\varepsilon}_{\bm{\omega}}$.
      \STATE Initialize $\bm{\varepsilon}_{\bm{\omega}}$.
      \FOR{each sampled mini-batch $\{(x^{\text{gen}}_i, x^{\text{adv}}_i)\}_{i=1}^B$}
          \FOR{each $(x^{\text{gen}}_i, x^{\text{adv}}_i)$}
            \STATE Generate $\pi$-noise $\bm{\varepsilon}_{\bm{\omega}}(x_i^{\mathrm{gen}})$ and $\bm{\varepsilon}_{\bm{\omega}}(x_i^{\mathrm{adv}})$ with Eq.~\eqref{eq:pnp-mix}.
            \STATE Obtain purified utterances $\hat{x}_i^{\mathrm{gen}}$ and $\hat{x}_i^{\mathrm{adv}}$ with Eq.~\eqref{eq:x_pnp}.
          \ENDFOR
         \STATE Compute $\mathcal{L}_{\mathrm{robust}}$ with the frozen ASV system with Eq.~\eqref{eq:robust-loss}, and compute $\mathcal{L}_{\mathrm{reg}}$ with Eq.~\eqref{eq:reg-loss}.
          \STATE Update $\bm{\varepsilon}_{\bm{\omega}}$ by minimizing $\mathcal{L}(\bm{\omega}) = \mathcal{L}_{\mathrm{robust}} + \gamma\mathcal{L}_{\mathrm{reg}}$.
      \ENDFOR
  \end{algorithmic}
\end{algorithm}

\begin{algorithm}[t]
  \caption{Purify a test sample with a trained positive-incentive noise predictor.}
  \label{alg:pnp-infer}
  \begin{algorithmic}[1]
      \REQUIRE Trained $\pi$-noise predictor $\bm{\varepsilon}_{\bm{\omega}}$ and test sample $x$.
      \ENSURE Purified sample $\hat{x}$.
      \STATE Predict the $\pi$-noise $\bm{\varepsilon}_{\bm{\omega}}(x)$ according to Eq.~\eqref{eq:pnp-mix}.
      \STATE Compute the purified sample $\hat{x}$ by Eq.~\eqref{eq:x_pnp}.
      \STATE Return $\hat{x}$ for downstream ASV scoring.
  \end{algorithmic}
\end{algorithm}

\subsection{Applications to Adversarial Purification}
\label{PnP-Diff+DM}

Based on the unified mixing formulation in Eq.~\eqref{eq:x_pnp}, PnP can be instantiated by choosing different weights for the input waveform and the predicted $\pi$-noise. We consider two basic standalone variants for ASV adversarial purification, namely PnP-Gaussian and PnP-Diff. In addition, PnP-Diff can be optionally cascaded with a diffusion denoiser as an extension for further processing.

\subsubsection{PnP-Gaussian}
\label{PnP-Gaussian}
When $w_x = w_n = 1$, PnP reduces to simply adding the learned noise to the original input:
\begin{equation}
\label{eq:forward_pnp_gau}
\hat{x} \;=\; x \;+\; \bm{\varepsilon}_{\bm{\omega}}(x).
\end{equation}
We refer to this simplest variant as PnP-Gaussian.

\subsubsection{PnP-Diff}
\label{PnP-Diff}
In vanilla diffusion models, the forward process is a Markov chain that gradually corrupts the input by combining the clean signal and Gaussian noise with coefficients determined by a noise schedule. Motivated by this formulation, we define a diffusion-style variant, PnP-Diff, which aligns PnP with the diffusion forward process. Specifically, we set the weights $w_x$ and $w_n$ according to the standard cumulative diffusion noise schedule $\{\bar{\alpha}_t\}$, where $\sqrt{\bar{\alpha}_t}$ and $\sqrt{1-\bar{\alpha}_t}$ weight the signal and noise terms, respectively. The PnP-Diff output at timestep $t$ is then given by
\begin{equation}
  \label{eq:forward-pnp-diff}
  \hat{x} \;=\; \sqrt{\bar{\alpha}_t}\,x \;+\; \sqrt{1-\bar{\alpha}_t}\,\bm{\varepsilon}_{\bm{\omega}}(x,t).
\end{equation}
Compared with PnP-Gaussian, PnP-Diff explicitly follows the diffusion noise schedule and can therefore be combined naturally with a diffusion reverse process starting from $\hat{x}$.

\begin{algorithm}[t]
  \caption{Train a diffusion model with PnP-generated $\pi$-noise}
  \label{alg:pnp-diffusion}
  \begin{algorithmic}[1]
      \REQUIRE Trained PnP-Diff generator $\bm{\varepsilon}_{\bm{\omega}}$, training utterances $\{x_i\}_{i=1}^n$, batch size $B$, maximum diffusion timestep $T$.
      \ENSURE Diffusion model $\bm{\epsilon}_{\bm{\theta}}$.
      \STATE Initialize $\bm{\epsilon}_{\bm{\theta}}$.
      \FOR{each sampled mini-batch $\{x_i\}_{i=1}^B$}
        \STATE Sample timesteps $t_i \sim \mathrm{Uniform}\{1,\dots,T\}$ for $i=1,\dots,B$.
        \FOR{each $x_i$ in the mini-batch}
          \STATE Generate $\pi$-noise $\bm{\varepsilon}_i = \bm{\varepsilon}_{\bm{\omega}}(x_i,t_i)$.
          \STATE Form the corrupted sample $x_{t,i}$ from $x_i$ and $\bm{\varepsilon}_i$ according to Eq.~\eqref{eq:forward-pnp-diff}.
        \ENDFOR
        \STATE Compute the diffusion loss $
    \frac{1}{B}\sum_{i=1}^{B}
    \big\|
      \bm{\varepsilon}_i
      - \bm{\epsilon}_{\bm{\theta}}(x_{t,i}, t_i)
    \big\|_2^2$.
        \STATE Update $\bm{\theta}$ by minimizing the diffusion loss.
      \ENDFOR
  \end{algorithmic}
\end{algorithm}

\subsubsection{Optional diffusion denoiser cascade}

PnP-Diff supports an optional extension in which its forward output is further processed by a diffusion denoiser. This denoiser can be either a pretrained model or a model retrained with PnP-generated $\pi$-noise.

For this cascade, we treat the PnP-Diff output as a diffusion-style corrupted sample, since it follows the same noise schedule as the standard diffusion forward process. In this case, as shown in Algorithm~\ref{alg:pnp-diffusion}, we use $x_t$ rather than $\hat{x}$ to denote the mixture of $x$ and $\bm{\varepsilon}_{\bm{\omega}}(x,t)$, so that it aligns better with the diffusion model. Specifically, PnP-Diff replaces Gaussian noise in the diffusion forward process with learned $\pi$-noise $\bm{\varepsilon}_{\bm{\omega}}(x,t)$, and constructs $x_t$ according to Eq.~\eqref{eq:forward-pnp-diff}. This allows a diffusion reverse process to be applied after the PnP-Diff forward step for further quality enhancement.

On top of this PnP-driven forward process, we can then retrain a diffusion denoiser $\bm{\epsilon}_{\bm{\theta}}(x_t,t)$ to predict the injected $\pi$-noise. The diffusion model loss is formulated as
\begin{equation}
  \label{eq:diff-loss}
  \mathcal{L}_{\mathrm{diff}}
  = \mathbb{E}_{x,\,t}\,
    \mathbb{E}_{\bm{\varepsilon}_{\bm{\omega}}(x,t)}
    \Big[
      \big\|
        \bm{\varepsilon}_{\bm{\omega}}(x,t)
        - \bm{\epsilon}_{\bm{\theta}}(x_t,t)
      \big\|_2^2
    \Big].
\end{equation} This loss function encourages the reverse process to remove the PnP-generated $\pi$-noise. Algorithm~\ref{alg:pnp-diffusion} summarizes the corresponding training procedure.

\section{Experimental Settings}
\label{sec:exp_settings}

\subsection{Datasets}
We conduct most of experiments on the VoxCeleb benchmarks \cite{nagrani2017voxceleb,chung2018voxceleb2}.
 VoxCeleb1 and VoxCeleb2 dataset are extracted from YouTube videos and include thousands of speakers. To train the ASV backbones, we follow common protocol and use the full VoxCeleb2 dataset, except pretrained SimAMResNet using VoxBlink2 \cite{lin2024voxblink2}. The VoxCeleb1 development subset of  is used to train the proposed PnP methods with VoxCeleb1-E trial list. 

We evaluate adversarial attack and defense performance over 4,000 paired verification trials from VoxCeleb1 test set. To balance computational cost and generalizability, we construct this subset by randomly sampling 50 non-target trials and 50 target trials for each enrollment speaker from the VoxCeleb1-O trial list. In addition, because there are no transcriptions for VoxCeleb dataset, 1,000 utterances are randomly selected from the LibriSpeech train-clean-100 subset  \cite{panayotov2015librispeech} and used for word error rate (WER) evaluation.

\subsection{ASV systems}
We conduct adversarial attack and defense experiments on four widely used state-of-the-art ASV backbones: ECAPA-TDNN~\cite{desplanques2020ecapa}, CAM++~\cite{wang2023cam++}, ResNet~\cite{zeinali2019but}, and SimAMResNet~\cite{qin2022simple}. All models are implemented using the WeSpeaker toolkit~\cite{wang2023wespeaker} and take 80-dimensional LogFBank coefficients as input features. 

ECAPA-TDNN and CAM++ are lightweight embedding extractors that produce 512-dimensional speaker embeddings, with 6.4M and 7.2M parameters, respectively. In contrast, ResNet and SimAMResNet provide higher-capacity backbones. ResNet has 221 layers and outputs 256-dimensional embeddings with 23.8M parameters, while SimAMResNet has 100 layers, also outputs 256-dimensional embeddings, and contains 50.2M parameters.

\subsection{Adversarial attacks}

For white-box attacks, we evaluate three strong gradient-based waveform-domain attacks: MI-FGSM~\cite{dong2018boosting}, PGD-$\ell_\infty$, and PGD-$\ell_2$~\cite{madry2018towards}. For MI-FGSM and PGD-$\ell_\infty$, we set the step size to $\alpha=1$, the $\ell_\infty$ perturbation budget to 30, and the momentum decay in MI-FGSM to $\mu=1$. For PGD-$\ell_2$, we set the step size to $\alpha=500$ and the perturbation budget to 6400. All white-box attacks are targeted and implemented with the torchattacks\footnote{\url{https://github.com/Harry24k/adversarial-attacks-pytorch}} toolkit. In addition, all white-box attacks are also evaluated under a defender-aware adaptive attack setting, where the attacker also has access to the architecture and parameters of purification methods.

For black-box attacks, we consider FAKEBOB~\cite{chen2021real}, a query-based targeted false-accept attack for ASV. We set the perturbation budget to $\bm\epsilon_{FB}=160$, and the maximum number of iterations to 150. Our implementation follows the public FAKEBOB\footnote{\url{https://github.com/FAKEBOB-adversarial-attack/FAKEBOB}} repository. 

All attacks are targeted because they correspond to ASV impersonation scenarios, where an attacker attempts to be accepted as a target speaker and gain unauthorized access.

\subsection{Purification methods}
We compare the proposed PnP with several representative purification baselines. 
DAP~\cite{bai2024diffusion} is a diffusion-based adversarial purification method for ASV that adopts a DiffWave~\cite{kong2020diffwave} backbone conditioned on Mel-spectrograms; we train it under the same adversarial data and setting as PnP-Diff for a fair comparison. 
AudioPure uses a diffusion model pretrained on genuine data; we use the official checkpoint\footnote{\url{https://github.com/cychomatica/AudioPure}} for testing.
We additionally implement AudioPure-SSNI following SSNI~\cite{sun2025sample}. We retain the pretrained AudioPure backbone and replace the fixed purification step with a sample-specific step determined from the predicted-noise trajectory.
We also include simple additive Gaussian-noise baselines~\cite{chang2021defending}, denoted as Noise-$\sigma$, where zero-mean Gaussian noise with standard deviation $\sigma\in\{0.005,0.01,0.02\}$ is directly added to the waveform. 
For neural codec-based purifiers \cite{chen2024neural}, the input waveform is compressed into a latent representation and then decompressed. We evaluate three pretrained codecs at 16\,kHz: SpeechTokenizer\footnote{\url{https://huggingface.co/OpenMOSS-Team/SpeechTokenizer}}~\cite{zhangspeechtokenizer}, DAC\footnote{\url{https://github.com/descriptinc/descript-audio-codec}}~\cite{kumar2023high}, and AcademiCodec (HiFi-Codec-16k-320d-large-universal)\footnote{\url{https://github.com/yangdongchao/AcademiCodec}}~\cite{yang2023hifi}.

For PnP, we instantiate an unconditional 1-D U-Net DiffWave~\cite{kong2020diffwave} as the noise predictor and consider two variants, \textbf{PnP-Gaussian} and \textbf{PnP-Diff}. PnP-Gaussian directly adds the predicted noise to the input, whereas PnP-Diff follows the diffusion-style forward corruption defined in Eq.~\eqref{eq:forward-pnp-diff}. For both variants, the mixture coefficient in Eq.~\eqref{eq:pnp-mix} is fixed to $\lambda=0.7$. For PnP-Diff, we use a 50-step linear noise schedule with $\beta_t$ linearly increasing from $10^{-4}$ to $0.05$. During training, PnP-Diff samples timesteps uniformly from $t\in\{1,2,3\}$, while PnP-Gaussian is timestep-free. At inference time, we use $t^*=1$ as the default purification step for all PnP-Diff- and AudioPure-based methods, since a smaller timestep introduces milder degradation to the input. Other timestep settings are indicated in the method name, e.g., PnP-Diff-2 denotes $t^*=2$.

PnP-Diff is trained on 50-step PGD-$\ell_2$ adversarial examples generated against ECAPA-TDNN using the VoxCeleb1 development data and the VoxCeleb1-E trial list, with the margin in Eq.~\eqref{eq:robust-loss} set to $m=0.9$. PnP-Gaussian is trained on the 20-step ECAPA PGD-$\ell_2$ adversarial set from the same trials with margin $m=1.0$.

In addition, PnP-Diff can be cascaded with a diffusion denoiser to further improve perceptual quality. In this setting, PnP-Diff serves as a replacement for the standard forward noising process. We evaluate two options, \textbf{PnP-Diff + AudioPure}, which cascades PnP-Diff with a pretrained AudioPure denoiser, and \textbf{PnP-Diff + DiffWavePnP}, which uses a DiffWave denoiser trained with $\pi$-noise generated from PnP-Diff on the same 50-step PGD-$\ell_2$ VoxCeleb1 development data.

\subsection{Evaluation Metrics}
We report three groups of metrics covering ASV defense performance, audio quality and intelligibility, and inference speed.

For ASV defense performance and detection, we measure the equal error rate (EER) of the ASV system before and after purification, reported separately on genuine (noattacked) inputs and adversarial data for each trial. We also evaluate adversarial example detection by thresholding the ASV score shift induced by purification and report the corresponding detection rate when distinguishing adversarial from genuine inputs.

For audio quality and intelligibility, we report objective metrics, including wideband perceptual evaluation of speech quality (WB-PESQ), scale-invariant signal-to-distortion ratio (SI-SDR), short-time objective intelligibility (STOI), MOS prediction from a finetuned self-supervised model (MOS-SSL)~\cite{cooper2022generalization}, and WER. WB-PESQ, SI-SDR, and STOI are computed using the torchmetrics\footnote{\url{https://github.com/Lightning-AI/torchmetrics}} toolkit. For MOS-SSL, we use a finetuned wav2vec 2.0 small model as the SSL predictor. WER is evaluated with a Whisper\footnote{\url{https://huggingface.co/Systran/faster-whisper-large-v3}} model.
For inference speed, we report the real-time factor (RTF) of each method, defined as the ratio between model processing time and input duration. ASV processing time is also included.

\section{Experimental Results}

We first show the effect of adversarial attacks on the victim ASV systems and then organize the discussion around three questions:
\textbf{(Q1)} Is the reverse process really necessary in diffusion-based purification?
\textbf{(Q2)} How effective is the proposed PnP method against black-box, white-box and adaptive adversarial attackers?
\textbf{(Q3)} How can PnP be cascaded with diffusion models to further improve the quality and intelligibility for purified utterances?

\subsection{Attack performance without purification}
\label{subsec:attacker_wo_puri}

Before evaluating purification, we first verify that the attack settings are sufficiently strong for unprotected ASV systems. For white-box attacks, we consider MI-FGSM, PGD-$\ell_2$, and PGD-$\ell_\infty$ as three targeted methods. For each ASV backend and each attack, we fix the perturbation budget and vary the iteration number $K\in\{5,10,20,50,100,200\}$ to control the attack strength. For genuine inputs, ECAPA-TDNN, CAM++, ResNet, and SimAMResNet achieve EERs of 1.25\%, 0.85\%, 0.70\%, and 0.20\%, respectively, indicating that the backbones themselves are reliable before adversarial attack.

\begin{figure*}[t]
  \centering
  \includegraphics[width=0.92\textwidth]{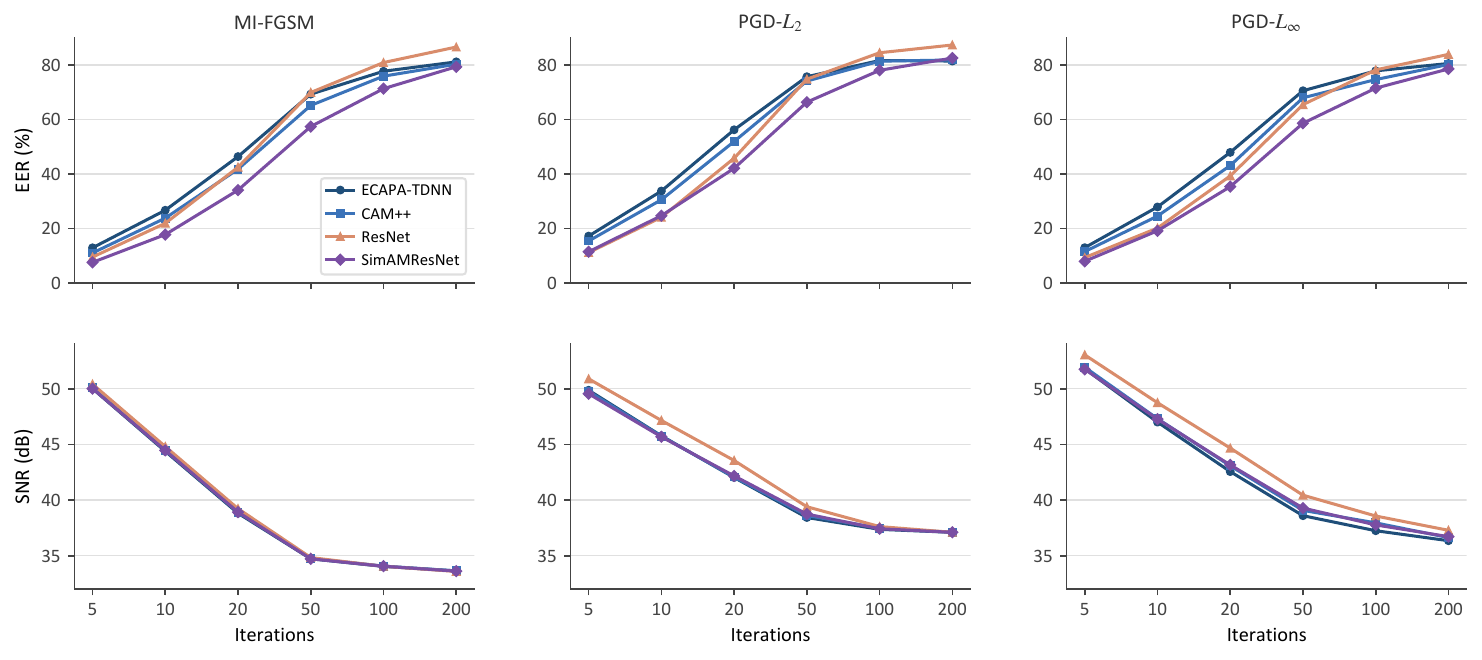}
  \caption{Attack strength on four unprotected ASV systems across attack iterations. The top row reports EER (\%) on attacked test trials and the bottom row reports average SNR (dB) for adversarial perturbations. The three columns correspond to MI-FGSM, PGD-$\ell_2$, and PGD-$\ell_\infty$, respectively, while different curves denote different ASV backbones.}
  \label{fig:attacker-strength}
\end{figure*}

As shown in Fig.~\ref{fig:attacker-strength}, increasing $K$ rapidly raises the EER for all three white-box attacks across all four ASV backbones. When the iteration number reaches $200$, EERs are already in the order of 80\% in most cases. At the same time, results in the bottom row show that perturbations remain, with the average SNR between unattacked and attacked waveforms consistently above 30 dB.

For the black-box FAKEBOB attack, we report attack success rate (ASR) rather than EER as this is a query-based false-accept attack. Without purification, FAKEBOB reaches 76.60\% ASR on the test subset, while the generated adversarial utterances have an average SNR of 24.95 dB. Overall, these results confirm that the unprotected ASV systems are highly vulnerable under both white-box and black-box attacks. Detailed EER values and minDCF results are provided in the supplementary material.

\subsection{Analysis of diffusion-based purification}
\label{subsec:audiopure}

To better understand what drives diffusion-based purification, we revisit AudioPure~\cite{wu2023defending} as a representative baseline. We study AudioPure on ECAPA-TDNN under 50-step MI-FGSM, where the unprotected system yields 1.25\% EER on genuine trials and 69.35\% EER on adversarial trials. AudioPure consists of a forward noising stage and a reverse denoising stage. In the forward process, a waveform $x$ is perturbed into $x_t=\sqrt{\bar{\alpha}_t}\,x+\sqrt{1-\bar{\alpha}_t}\,\bm{\epsilon}$, where $\bm{\epsilon}$ is injected Gaussian noise and $\bar{\alpha}_t$ is the cumulative noise schedule.

We compare three variants: (i) \emph{Full}, which uses both forward noising and reverse denoising; (ii) \emph{Forward}, which uses only the forward noising stage; and (iii) \emph{Forward without noise }, which uses only $x_t=\sqrt{\bar{\alpha}_t}\,x$.

Table~\ref{tab:stepwise-diff} reports the EERs and RTFs of these variants for different purification timesteps $t^*$. In terms of post-purification EERs on genuine and adversarial trials, forward-only AudioPure remains close to the full pipeline across all tested $t^*$, whereas its no-noise variant yields EERs similar to those without purification. This answers \textbf{Q1} and suggests that injected Gaussian noise in diffusion-based purification is the main source of robustness. These observations motivate PnP, which removes the costly reverse process and learns input-adaptive $\pi$-noise.

\subsection{Purification performance}
\label{subsec:asv-impact}

\begin{table}[t]
  \caption{Analysis of three AudioPure variants on ECAPA-TDNN under 50-step MI-FGSM attack. ``Gen'' and ``Adv'' denote post-purification EERs on genuine and adversarial trials, respectively. The no-purification EERs are 1.25\% for ``Gen'' and 69.35\% for ``Adv''. RTF is measured on a 24-GB NVIDIA GeForce RTX 3090 GPU.}
  \label{tab:stepwise-diff}
  \centering
  \setlength{\tabcolsep}{4pt}
  \renewcommand{\arraystretch}{1.05}
  \begin{tabular}{c ccc ccc ccc}
    \toprule
    \multicolumn{10}{c}{\textbf{AudioPure}} \\
    \midrule
    \multirow{2}{*}{$t^*$} &
    \multicolumn{3}{c}{Full} &
    \multicolumn{3}{c}{Forward} &
    \multicolumn{3}{c}{Forward without noise} \\
    \cmidrule(lr){2-4}\cmidrule(lr){5-7}\cmidrule(lr){8-10}
     & Gen& Adv & RTF & Gen& Adv & RTF & Gen& Adv & RTF \\
    \midrule
    $1$ & 2.90& 3.85& 0.078 & 2.80 & 4.05& 0.007 & 1.25 & 69.35& 0.007 \\
    $2$ & 4.35 & 4.70& 0.133 & 4.15 & 4.90& 0.007 & 1.25 & 69.35& 0.007 \\
    $3$ & 6.40& 7.00& 0.212 & 5.70 & 5.95& 0.007 & 1.25 & 69.35& 0.007 \\
    \bottomrule
  \end{tabular}
 \end{table}

\begin{table*}[t]
  \caption{Impact of different purification methods on ECAPA-TDNN for genuine unattacked input, white-box and black-box attacks, and inference speed. The genuine column reports EER (\%) on unattacked trials after applying each defender. The white-box columns report EER (\%) under targeted MI-FGSM, PGD-$\ell_2$, and PGD-$\ell_\infty$ attacks, and `Pooled' denotes the mean EER over iterations $\{10,50,200\}$ for each attacker. The FAKEBOB column reports black-box attack success rate. RTF measures the inference speed of each defender on a 48-GB NVIDIA L40S GPU.}
  \label{tab:whitebox-ecapa-10-50-200-pooled-per-attacker}
  \centering
  \setlength{\tabcolsep}{3.8pt}
  \renewcommand{\arraystretch}{1.05}
  \begin{tabular}{
    l
    !{\vrule}
    c
    !{\vrule}
    *{4}{c}
    !{\vrule}
    *{4}{c}
    !{\vrule}
    *{4}{c}
    !{\vrule}
    c
    !{\vrule}
    c
  }
    \toprule
\multicolumn{1}{l}{\multirow{3}{*}{Method}} &
\multicolumn{1}{l}{\hspace{-7pt}\multirow{3}{*}{Genuine}} &
\multicolumn{12}{c}{White-box} &
\multicolumn{1}{c@{\hspace{8pt}}}{Black-box} &
\multicolumn{1}{c}{\multirow{3}{*}{RTF}} \\
\cmidrule(lr){3-14}\cmidrule(lr){15-15}

\multicolumn{1}{c}{} &
\multicolumn{1}{c}{} &
\multicolumn{4}{c}{MI-FGSM} &
\multicolumn{4}{c}{PGD-$\ell_2$} &
\multicolumn{4}{c}{PGD-$\ell_\infty$} &
\multicolumn{1}{c@{\hspace{8pt}}}{FAKEBOB} &
\multicolumn{1}{c}{} \\
\cmidrule(lr){3-6}\cmidrule(lr){7-10}\cmidrule(lr){11-14}

\multicolumn{1}{c}{} &
\multicolumn{1}{c}{} &
\multicolumn{1}{c}{10} &
\multicolumn{1}{c}{50} &
\multicolumn{1}{c}{200} &
\multicolumn{1}{c}{Pooled} &
\multicolumn{1}{c}{10} &
\multicolumn{1}{c}{50} &
\multicolumn{1}{c}{200} &
\multicolumn{1}{c}{Pooled} &
\multicolumn{1}{c}{10} &
\multicolumn{1}{c}{50} &
\multicolumn{1}{c}{200} &
\multicolumn{1}{c}{Pooled} &
\multicolumn{1}{c@{\hspace{8pt}}}{ASR (\%)} &
\multicolumn{1}{c}{} \\
\midrule

    No defender
    & 1.25
    & 26.70 & 69.35 & 81.20 & 59.08
    & 33.75 & 75.75 & 81.50 & 63.67
    & 27.90 & 70.60 & 80.60 & 59.70
    & 76.60
    & \multicolumn{1}{c}{--} \\
\midrule

    \multicolumn{16}{l}{\textit{Diffusion-based baselines}} \\
    DAP~\cite{bai2024diffusion}, 2024
    & \textbf{1.65}
    & 9.30 & 28.15 & 20.30 & 19.25
    & 9.40 & 21.85 & 21.90 & 17.72
    & 6.35 & 15.10 & 17.20 & 12.88
    & 6.40
    & 0.072 \\
    AudioPure~\cite{wu2023defending}, 2023
    & 2.90
    & 3.05 & 3.85 & 3.85 & 3.58
    & 3.15 & 3.55 & 3.75 & 3.48
    & 2.95 & 3.75 & 3.55 & 3.42
    & 2.87
    & 0.050 \\
    AudioPure-SSNI~\cite{sun2025sample}, 2025
    & 2.95
    & 3.45 & 4.05 & 4.05 & 3.85
    & 3.50 & 3.70 & 3.70 & 3.63
    & 3.40 & 3.75 & 3.90 & 3.68
    & 3.20
    & 1.085 \\
    \midrule

    \multicolumn{16}{l}{\textit{Additive noise baselines}} \\
    Noise-0.005
    & 2.20
    & 2.70 & 4.75 & 4.40 & 3.95
    & 2.80 & 3.80 & 3.85 & 3.48
    & 2.75 & 3.75 & 3.70 & 3.40
    & 4.07
    & \textbf{0.006} \\
    Noise-0.01
    & 2.95
    & 3.15 & \textbf{3.80} & 3.95 & 3.63
    & 3.20 & \textbf{3.35} & 3.65 & 3.40
    & 3.10 & 3.60 & 3.60 & 3.43
    & 3.13
    & \textbf{0.006} \\
    Noise-0.02~\cite{chang2021defending}, 2021
    & 4.35
    & 5.10 & 5.30 & 5.00 & 5.13
    & 4.85 & 5.00 & 5.15 & 5.00
    & 4.55 & 5.15 & 5.15 & 4.95
    & 2.53
    & \textbf{0.006} \\
    \midrule

    \multicolumn{16}{l}{\textit{Neural codec baselines}} \\
    SpeechTokenizer~\cite{zhangspeechtokenizer}, 2024
    & 2.50
    & 5.75 & 9.85 & 7.70 & 7.77
    & 5.70 & 8.65 & 8.55 & 7.63
    & 4.90 & 7.40 & 7.75 & 6.68
    & 5.67
    & 0.018 \\
    DAC~\cite{kumar2023high}, 2023
    & 1.70
    & 7.85 & 18.85 & 14.45 & 13.72
    & 7.75 & 16.45 & 16.35 & 13.52
    & 5.70 & 13.45 & 14.20 & 11.12
    & 3.33
    & 0.017 \\
    AcademiCodec~\cite{yang2023hifi}, 2023
    & 5.35
    & 6.95 & 8.65 & 7.85 & 7.82
    & 7.10 & 8.25 &  8.50  & 7.95
    & 6.65 & 8.30 & 8.25 & 7.73
    & \textbf{2.40}
    & 0.017 \\
    \midrule

    \multicolumn{16}{l}{\textit{Proposed PnP (ours)}} \\
    PnP-Gaussian
    & 4.30
    & 5.30 & 6.70 & 5.65 & 5.88
    & 5.15 & 6.15 & 5.95 & 5.75
    & 5.15 & 5.75 & 5.40 & 5.43
    & 3.07
    & 0.014 \\
    PnP-Diff
    & 1.75
    & 2.65 & 4.45 & 3.65 & 3.58
    & \textbf{2.60} & 3.55 & \textbf{3.60} & \textbf{3.25}
    & \textbf{2.35} & 3.50 & \textbf{3.20} & \textbf{3.02}
    & 3.40
    & 0.014 \\
   PnP-Diff-2
   & 2.40
   & 2.90 & 4.10 & \textbf{3.50} & \textbf{3.50}
   & 3.00 & 3.40 & 3.85 & 3.42
   & 2.75 & \textbf{3.35} & 3.50 & 3.20
   & \textbf{2.40}
   & 0.014 \\
    PnP-Diff + AudioPure
    & 1.80
    & \textbf{2.60} & 4.35 & 3.70 & 3.55
    & 2.65 & 3.90 & \textbf{3.60} & 3.38
    & \textbf{2.35} & 3.45 & 3.30 & 3.03
    & 3.07
    & 0.091 \\
    PnP-Diff + DiffWavePnP
    & 2.35
    & 3.30 & 5.45 & 5.20 & 4.65
    & 2.65 & 4.80 & 4.85 & 4.10
    & 3.45 & 4.55 & 4.35 & 4.12
    & 4.27
    & 0.017 \\
    \bottomrule
  \end{tabular}
\end{table*}

We next use ECAPA-TDNN as a representative ASV backbone to evaluate the end-to-end impact of purification. Results of PnP on other ASV systems are reported in Section~\ref{sec:abl_ASVs}.
 Table~\ref{tab:whitebox-ecapa-10-50-200-pooled-per-attacker} illustrates the influence of purification on genuine unattacked inputs, white-box and black-box attacks, and inference speed for all defenders. The "No defender" row reports EERs without purification and serves as the baseline to measure the reductions achieved by each purifier.

For genuine unattacked inputs, DAP has the lowest impact, but the proposed PnP-Diff variants introduce only a small degradation. Compared with the no-purification EER of 1.25\%, DAP and PnP-Diff obtain 1.65\% and 1.75\%, respectively. In contrast, PnP-Gaussian reaches 4.30\%, and the additive-noise baselines are all above 2\%, showing that the additive formulation is more harmful on clean trials. This confirms that the diffusion-style PnP design better preserves genuine ASV performance.

For white-box attacks, adversarial examples are generated against ECAPA-TDNN and then evaluated on the same ASV after purification. Among all methods, PnP-Diff provides the most stable trade-off on genuine and attacked data. It maintains low EERs for genuine utterances while achieving the strongest pooled white-box EERs across PGD-$L_2$ and PGD-$L_\infty$. Here, PnP-Diff-2 denotes the two-step forward purification version of PnP-Diff. The two cascaded variants further show that the main robustness gain already comes from the learned forward purification.

The black-box FAKEBOB column tests whether this trend transfers beyond gradient-based attacks. Since FAKEBOB targets false accepts at a fixed verification threshold, we report attack success rate (ASR) rather than EER. PnP-Diff-2 and AcademiCodec achieve the best results, reducing the no-purification ASR from 76.60\% to 2.40\%. Overall, among all defenders, the 1-step PnP-Diff and its 2-step variant, PnP-Diff-2, provide the strongest performance under both white-box and black-box attacks, and cause only limited degradation on genuine inputs.

RTF evaluation further highlights PnP's inference-speed advantage. PnP-Gaussian, PnP-Diff, and PnP-Diff-2 all run at an RTF of 0.014, which is much faster than reverse-diffusion pipelines such as AudioPure. The cascaded PnP-Diff + AudioPure and PnP-Diff + DiffWavePnP variants are slower because they additionally invoke a diffusion denoiser. Overall, the proposed PnP variants provide a favorable balance among low genuine-input degradation, strong white-box and black-box robustness, and inference efficiency.

\subsection{Adaptive attack robustness}
\label{subsec:adaptive-attack}

After showing that the PnP framework improves adversarial robustness under standard attacks, we next evaluate defender-aware adaptive attacks and answer \textbf{Q2}. This experiment tests the claim that PnP should remain useful even when the attacker has full knowledge of the purifier architecture and
parameters. We therefore assume that the attacker computes gradients through the whole defended ASV pipeline and evaluate the resulting EER under this adaptive attack setting.

\begin{figure}[t]
  \centering
  \includegraphics[width=0.9\columnwidth]{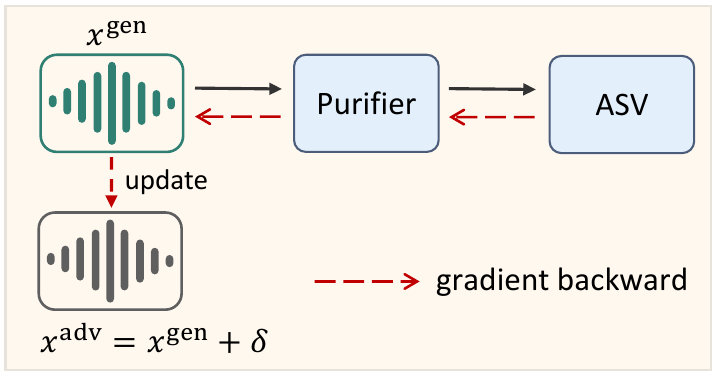}
  \caption{Adaptive attack against the defended ASV pipeline. Black arrows denote the defended forward inference from the clean input $x^{\mathrm{gen}}$ through the purifier and the ASV system. Red dashed arrows denote the backward gradient path used by the attacker to update the adversarial input $x^{\mathrm{adv}} = x^{\mathrm{gen}} + \delta$ through the whole defended ASV pipeline.}
  \label{fig:adaptive-pipeline}
\end{figure}

As illustrated in Fig.~\ref{fig:adaptive-pipeline}, the attack objective is defined on the defended pipeline $\text{ASV}(\text{Purifier}(x))$ rather than only on the verifier $\text{ASV}(x)$. To keep the comparison focused, from this point we report only the best defender from each baseline family according to Table~\ref{tab:whitebox-ecapa-10-50-200-pooled-per-attacker}.

\begin{table}[t]
  \caption{Adaptive 50-step MI-FGSM attack evaluation on ECAPA-TDNN. We report the EER of adaptive examples evaluated with the same attacked defender in the pipeline, together with the EER of the same examples evaluated by the ASV alone.}
  \label{tab:adaptive-attack}
  \centering
  \setlength{\tabcolsep}{4pt}
  \renewcommand{\arraystretch}{1.05}
  \small
\begin{tabular}{lcc}
\toprule
& \multicolumn{2}{c}{EER on adaptive examples} \\
\cmidrule(lr){2-3}
Attacked Defender
& with defender
& without defender \\
\midrule
    No defender & -- & 69.35 \\
    AudioPure & 10.40 & 8.60 \\
    Noise-0.01 & 41.60 & 2.40 \\
    PnP-Gaussian & 53.30 & 1.20 \\
    PnP-Diff & 11.15 & 14.60 \\
    PnP-Diff-2 & \textbf{9.35} & 2.70 \\
    PnP-Diff + AudioPure & 11.95 & 14.60 \\
    PnP-Diff + DiffWavePnP & 20.60 & 5.10 \\
    \bottomrule
  \end{tabular}
\end{table}

Table~\ref{tab:adaptive-attack} focuses on 50-step MI-FGSM attacks and reports the EER of adaptive examples when the attacked defender remains in the ASV pipeline, together with the EER of the same examples after removing the defender. Among the PnP variants, PnP-Diff-2 is the most stable under the matched adaptive evaluation. In contrast, PnP-Gaussian is much more vulnerable. However, after removing the defender, its adaptive examples no longer remain effective on the bare ASV again.

\subsection{Adversarial example detection performance}
\label{subsec:detection}

\begin{table}[t]
  \caption{Adversarial example detection rate (DR) at FPR=1\% with the absolute ASV score difference before and after purification. The 50-step PGD-$\ell_2$ EER is shown as a reference for purification performance under the same attack setting. *MDD result is from the original paper with the same attack.}
  \label{tab:detection}
  \centering
  \setlength{\tabcolsep}{4pt}
  \renewcommand{\arraystretch}{1.05}
  \begin{tabular}{l c c}
    \toprule
    Method & DR(\%)@FPR=1\% $\uparrow$ & PGD EER (\%) $\downarrow$ \\
    \midrule
    MDD-10\%*& 94.10 & 18.00 \\
    AudioPure & 94.75 & 3.55 \\
    Noise-0.01 & 94.55 & \textbf{3.35} \\
    SpeechTokenizer & 95.95 & 8.65 \\
    PnP-Gaussian & 93.40 & 6.15 \\
    PnP-Diff & \textbf{96.60} & 3.55 \\
    PnP-Diff-2 & 95.60 & 3.40 \\
    PnP-Diff + AudioPure & \textbf{96.60} & 3.90 \\
    PnP-Diff + DiffWavePnP & 94.70 & 4.80 \\
    \bottomrule
  \end{tabular}
\end{table}

We next investigate whether our purification methods can also be used for adversarial example detection. In this task, an input utterance is classified as adversarial or genuine according to the ASV score difference before and after purification. If this difference is large, the input is more likely to be adversarial. By setting a target false positive rate (FPR), different decision thresholds can be obtained for different purifiers.

Table~\ref{tab:detection} shows that this detection strategy works well for most purifiers. The thresholds are computed on clean trials only and then applied to 50-step PGD-$\ell_2$ adversarial trials for detection. The best detection rate is achieved by PnP-Diff and PnP-Diff + AudioPure, both achieving 96.60\% at FPR=1\%. PnP-Gaussian gives the lowest detection rate among the methods at 93.40\%. This again shows that diffusion-style PnP variants provide a more stable detection cue than the additive variant.

It is also useful to evaluate the detection and purification performance jointly. Although Noise-0.01 achieves the lowest 50-step PGD-$\ell_2$ EER of 3.35\%, its detection rate is lower than those of the diffusion-style methods. PnP-Diff achieves the highest detection rate while maintaining a low EER of 3.55\%. We also compare with the previous MDD method with a 10\% mask ratio \cite{bai2025mdd}. Since MDD is specifically designed for adversarial example detection rather than purification, its corresponding PGD-$\ell_2$ EER remains as high as 18\%. By contrast, our method achieves strong performance on both purification and detection.

\subsection{Impact on input: audio quality and intelligibility analysis}
\label{subsec:audio-quality}

\begin{figure}[t]
\centering
\includegraphics[width=0.85\columnwidth]{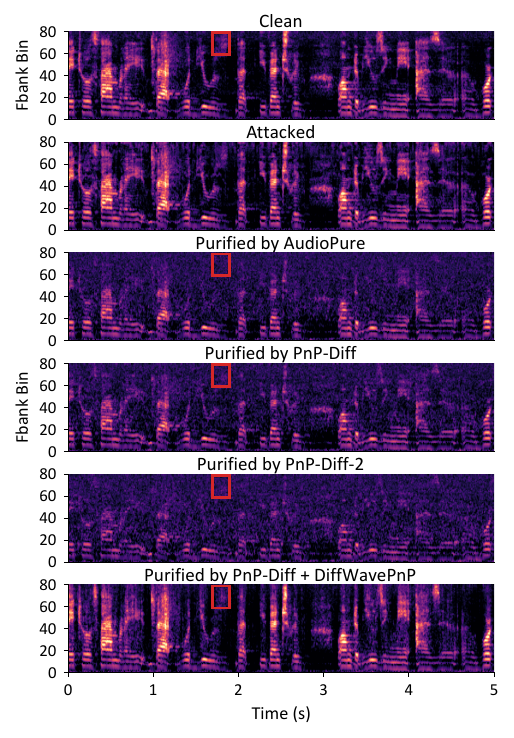}
\caption{Fbank feature comparison with selected purification methods under 50-step MI-FGSM. We show the clean input and the attacked version for reference, together with the outputs after applying different purification methods to the attacked input. The red boxes highlight an example high-frequency component. This component almost disappears after applying other purification methods, whereas PnP-Diff + DiffWavePnP preserves more of its detail.}
\label{fig:quality-case}
\end{figure}

In this section, we also examine how purification alters the input waveform itself. Table~\ref{tab:vox-quality} and Table~\ref{tab:libri-quality} report objective speech quality and intelligibility metrics, including WB-PESQ, SI-SDR, STOI, MOS-SSL, and WER, for selected waveform-domain defenses. Table~\ref{tab:vox-quality} reports the purification performance under 50-step MI-FGSM attack on ECAPA-TDNN. We use Whisper to compute the WER between purified samples and clean utterances. In Table~\ref{tab:libri-quality}, we additionally include LibriSpeech as a clean-speech benchmark to evaluate how strongly each method distorts speech content.

Both tables show a clear trade-off between audio quality and ASV robustness. Although PnP-Gaussian improves robustness, it also causes the largest degradation in waveform fidelity and intelligibility on both datasets. In contrast, the diffusion-style PnP variants preserve fidelity much better while maintaining competitive robustness. In Table~\ref{tab:vox-quality}, PnP-Diff reaches an EER of 4.45\% for adversarial inputs with substantially better SI-SDR, STOI, and WER than PnP-Gaussian, which shows that the diffusion-style formulation is important not only for robustness but also for preserving fidelity.

The cascaded variants further clarify the role of denoising after forward purification. For all metrics, PnP-Diff + AudioPure performs similarly to PnP-Diff on both datasets, suggesting that the pretrained AudioPure denoiser provides only limited enhancement because it is trained on Gaussian noise rather than on the learned $\pi$-noise used by PnP-Diff. By contrast, PnP-Diff + DiffWavePnP delivers the best waveform-level fidelity among the defenders, achieving the strongest WB-PESQ, SI-SDR, and STOI results on both datasets. It also improves MOS-SSL over AudioPure and PnP-Diff on both datasets. This answers \textbf{Q3} and indicates that the cascaded DiffWave denoiser mainly serves as an extension for quality enhancement.

Fig.~\ref{fig:quality-case} provides a visual view of the Fbank features purified by selected diffusion-based methods. Compared with AudioPure, PnP-Diff produces similar features, and PnP-Diff + DiffWavePnP yields the cleanest restoration among the compared methods. The red-boxed region highlights a high-frequency component that is almost removed by the other purification methods but is more clearly preserved by PnP-Diff + DiffWavePnP. This suggests that the learned forward purification is the main source of robustness, while the optional denoising cascade can further improve perceptual fidelity and intelligibility. A full version of this visualization with all defenders in Table~\ref{tab:whitebox-ecapa-10-50-200-pooled-per-attacker} is provided in the supplementary material.

\begin{table*}[t]
  \caption{Objective speech quality and intelligibility evaluation on adversarial VoxCeleb data. MOS-SSL is predicted by a finetuned wav2vec~2.0 small model. The last column reports the EER of purified adversarial samples.}
  \label{tab:vox-quality}
  \centering
  \setlength{\tabcolsep}{3pt}
  \renewcommand{\arraystretch}{1.05}
  \begin{tabular}{lcccccc}
    \toprule
    Method & WB-PESQ $\uparrow$ & SI-SDR (dB) $\uparrow$ & STOI $\uparrow$ & MOS-SSL $\uparrow$& WER (\%) $\downarrow$ & Adv. EER (\%) $\downarrow$ \\
    \midrule
    No defender                        & 3.775& 34.70& 0.986& 2.337& 4.69& 69.35 \\
    \midrule
    AudioPure                          & 1.442  & 12.49   & 0.871  & 1.119 & 8.64  & 3.85 \\
    Noise-0.01                         & 1.437  & 12.42   & 0.871  & 1.113 & 8.49  & \textbf{3.80} \\
    SpeechTokenizer                    & 1.986  & -1.35   & 0.847  & 2.049 & 10.06 & 9.85 \\
    PnP-Gaussian                       & 1.057  & -9.45   & 0.539  & 0.403 & 23.69 & 6.70 \\
    PnP-Diff                           & 1.609  & 13.09   & 0.904  & 1.269 & 6.15  & 4.45 \\
    PnP-Diff-2                         & 1.172  & 3.67    & 0.775  & 0.673 & 9.61  & 4.10 \\
    PnP-Diff + AudioPure               & 1.616  & 13.15   & 0.905  & 1.275 & 5.94  & 4.35 \\
    PnP-Diff + DiffWavePnP             & \textbf{3.591}  & \textbf{21.14} & \textbf{0.958} & \textbf{2.065} & \textbf{5.80}  & 5.45 \\
    \bottomrule
  \end{tabular}
\end{table*}

\begin{table*}[t]
  \caption{Objective speech quality and intelligibility evaluation on clean LibriSpeech data. MOS-SSL is predicted by a finetuned wav2vec~2.0 small model.}
  \label{tab:libri-quality}
  \centering
  \setlength{\tabcolsep}{3pt}
  \renewcommand{\arraystretch}{1.05}
  \begin{tabular}{lccccc}
    \toprule
    Method & WB-PESQ $\uparrow$ & SI-SDR (dB) $\uparrow$ & STOI $\uparrow$ & MOS-SSL $\uparrow$ & WER (\%) $\downarrow$ \\
    \midrule
    No defender                        & --    & --     & --    & 4.180 & 3.88 \\
    \midrule
    AudioPure                          & 1.296 & 14.93  & 0.925 & 2.011 & 3.20 \\
    Noise-0.01                         & 1.292 & 14.87  & 0.925 & 1.999 & 3.20 \\
    SpeechTokenizer                    & 2.585 & 1.42   & 0.914 & \textbf{3.885} & 3.10 \\
    PnP-Gaussian                       & 1.042 & -10.14 & 0.612 & 0.474 & 6.58 \\
    PnP-Diff                           & 1.431 & 15.17  & 0.943 & 2.342 & 3.14 \\
    PnP-Diff-2                         & 1.096 & 5.08   & 0.847 & 1.008 & 3.23 \\
    PnP-Diff + AudioPure               & 1.437 & 15.23  & 0.944 & 2.354 & \textbf{3.02} \\
    PnP-Diff + DiffWavePnP             & \textbf{3.267} & \textbf{23.32} & \textbf{0.971} & 3.634 & 4.83 \\
    \bottomrule
  \end{tabular}
\end{table*}

\subsection{Ablation study}
\label{subsec:ablation}

Finally, we perform ablation studies to understand which components of the unified PnP framework are responsible for the observed robustness gains. We first analyze the main hyperparameters, then examine the effect of the purification step, and finally test how PnP-Diff performs across different ASV backbones.

\subsubsection{Effect of the main hyperparameters}

\begin{figure}[t]
  \centering
  \includegraphics[width=\columnwidth]{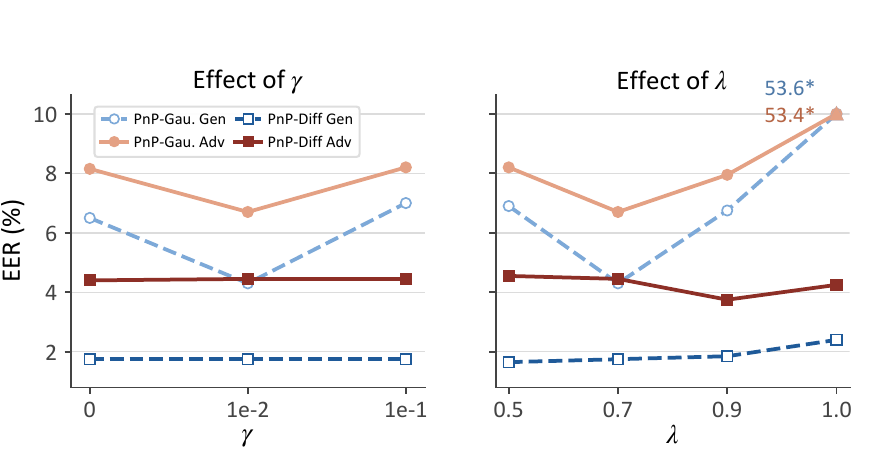}
  \caption{Ablation study on the main hyperparameters for PnP-Gaussian (PnP-Gau.) and PnP-Diff. Solid lines denote adversarial EER and dashed lines denote genuine EER after purification. * Values are outside the displayed y-range.}
  \label{fig:ablation-overview}
\end{figure}

We first examine the effects of the regularization weight $\gamma$ and the mixing factor $\lambda$ on PnP-Diff and PnP-Gaussian. For both variants, we report the purified EER on genuine unattacked inputs and adversarial inputs. As shown in Fig.~\ref{fig:ablation-overview}, PnP-Diff is less sensitive to both $\gamma$ and $\lambda$ than PnP-Gaussian. In the $\gamma$ study, $\gamma=10^{-2}$ provides a practical regularization trade-off by improving the stability of PnP-Gaussian without hurting the diffusion-style variant. In the $\lambda$ study, PnP-Diff remains within a relatively narrow EER range, whereas PnP-Gaussian performs best at $\lambda=0.7$ and degrades to random noising when $\lambda=1$. Overall, $\lambda=0.7$ or $0.9$ and $\gamma=10^{-2}$ provide a good trade-off.

\subsubsection{Effect of purification}
\label{subsec:step-ablation}

\begin{figure*}[t]
  \centering
  \includegraphics[width=0.92\textwidth]{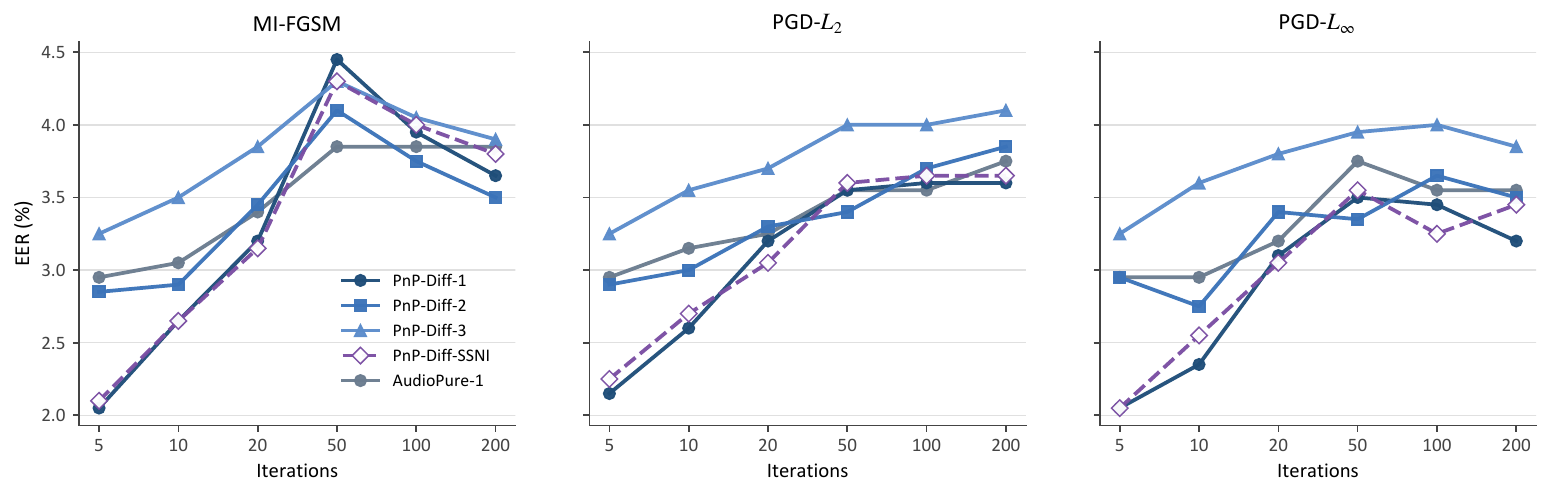}
  \caption{Effect of purification step on ECAPA-TDNN against three white-box attacks. We compare 1-step AudioPure with the 1-, 2-, and 3-step PnP-Diff, and the sample-specific variant PnP-Diff-SSNI. PnP-Diff-1 denotes the default PnP-Diff setting used in other tables. The x-axis shows the attack iteration number, and the y-axis reports the resulting EER (\%) after purification.}
  \label{fig:step-ablation-main}
\end{figure*}

We next examine how the purification step in PnP-Diff affects robustness. Fig.~\ref{fig:step-ablation-main} compares the 1-step AudioPure result with the 1-, 2-, and 3-step versions of PnP-Diff, as well as the sample-specific variant PnP-Diff-SSNI. The SSNI method~\cite{sun2025sample} adaptively selects the purification step for each sample. We evaluate different purification steps under three white-box attacks with different strengths. The full numerical EER and RTF results are provided in the supplementary material.

Fig.~\ref{fig:step-ablation-main} shows that small purification steps are already sufficient for PnP-Diff. Across MI-FGSM, PGD-$\ell_2$, and PGD-$\ell_\infty$ attacks, the 1-step and 2-step variants cover most of the useful robustness range, while the 3-step variant is usually worse. PnP-Diff-SSNI is competitive and occasionally gives the best result, but it does not provide a consistent improvement over the fixed 1-step or 2-step settings. These results suggest that a small step is already effective.

\subsubsection{Effect across ASV backbones}
\label{sec:abl_ASVs}
\begin{table}[t]
  \caption{PnP-Diff purification results with EER (\%) evaluated on four ASV architectures under 50-step MI-FGSM attack. Adversarial examples are generated separately for each ASV. ``Gen'' and ``Adv'' denote genuine and adversarial EER after purification.}
  \label{tab:ablation-architecture}
  \centering
  \setlength{\tabcolsep}{5pt}
  \renewcommand{\arraystretch}{1.05}
  \begin{tabular}{lcc}
    \toprule
    ASV architecture & Gen & Adv \\
    \midrule
    ECAPA-TDNN & 1.75 & 4.45 \\
    CAM++ & 1.20 & 2.90 \\
    ResNet & 1.00 & 0.70 \\
    SimAMResNet & 0.35 & 0.70 \\
    \bottomrule
  \end{tabular}
\end{table}

As shown in Table~\ref{tab:ablation-architecture}, we also evaluate PnP-Diff on different ASV backbones to defend against 50-step MI-FGSM attacks. Although it is trained on adversarial examples generated from ECAPA-TDNN, it still provides clear defensive gains for CAM++, ResNet, and SimAMResNet. This suggests that the learned $\pi$-noise transfers well across different ASV architectures. In particular, for ResNet, the EER for purified adversarial inputs is reduced to 0.70\%, which is even lower than the EER for genuine inputs after purification.

\section{Conclusion}

This paper proposes the Positive-Incentive Noise Predictor (PnP) framework, a simplification of previous noise-based and diffusion-based audio adversarial purification methods. Our analytics show that most robustness gains come from the forward noising stage. Based on this observation, we introduce positive-incentive theory to learn an input-adaptive noising process as a front-end purifier. Within this framework, we study two variants: PnP-Gaussian, which serves as the simplest learned additive-noising baseline, and PnP-Diff, which aligns the learned $\pi$-noise with the diffusion forward noise schedule. By combining PnP-Diff with a diffusion denoiser, we also provide a diffusion model trained with learned $\pi$-noise for quality enhancement.

The experimental results demonstrate the effectiveness of the proposed PnP methods. We compare the proposed methods with representative diffusion-based, additive-noise, and codec-based purification baselines, and test all of them under white-box and black-box attack settings. Under standard white-box attacks, the proposed PnP methods substantially reduce EERs for adversarial inputs, and PnP-Diff provides the strongest overall balance between EERs for purified genuine inputs, adversarial robustness, and inference efficiency. In addition, PnP-Diff remains effective across four ASV backbones, while PnP-Diff-2 is more stable under a 50-step defender-aware adaptive MI-FGSM attack. Audio quality analysis further shows that diffusion-style variants preserve speech quality better than the simple additive variant, and that an optional diffusion cascade can further improve waveform fidelity when perceptual quality is prioritized. In addition, PnP-Diff also performs well on the adversarial example detection task. These results indicate that learned forward noising is an appealing alternative to full reverse diffusion for adversarial purification.

\bibliographystyle{IEEEtran}
\bibliography{ref}
\clearpage
\onecolumn

\begin{center}
{\Large Supplementary Material}\\[0.4em]
{\normalsize Positive-Incentive Noise Predictor for Adversarial Purification in Speaker Verification}
\end{center}
\vspace{1em}

\begin{center}
  \captionof{table}{Attacker-strength results for unprotected ASV systems. We report EER (\%) and minDCF at different white-box attack iteration numbers.}
  \label{tab:attacker-strength-mindcf-supp}
  \setlength{\tabcolsep}{4pt}
  \renewcommand{\arraystretch}{1.05}
  \begin{tabular}{l c cc cc cc cc}
    \toprule
    \multirow{2}{*}{Attack} & \multirow{2}{*}{Iterations} &
    \multicolumn{2}{c}{ECAPA-TDNN} &
    \multicolumn{2}{c}{CAM++} &
    \multicolumn{2}{c}{ResNet} &
    \multicolumn{2}{c}{SimAMResNet} \\
    \cmidrule(lr){3-4}\cmidrule(lr){5-6}\cmidrule(lr){7-8}\cmidrule(lr){9-10}
     & & EER & minDCF & EER & minDCF & EER & minDCF & EER & minDCF \\
    \midrule
    No attacker  & --  & 1.25 & 0.101 & 0.85 & 0.107 & 0.70 & 0.069 & 0.20 & 0.011 \\
    \midrule
    \multirow{6}{*}{MI-FGSM}
      & 5   & 12.90 & 0.996 & 11.00 & 0.993 & 9.50 & 0.937 & 7.55 & 0.963 \\
      & 10  & 26.70 & 1.000 & 23.80 & 1.000 & 21.90 & 1.000 & 17.75 & 1.000 \\
      & 20  & 46.40 & 1.000 & 41.80 & 1.000 & 42.50 & 1.000 & 34.10 & 1.000 \\
      & 50  & 69.35 & 1.000 & 65.20 & 1.000 & 69.95 & 1.000 & 57.45 & 1.000 \\
      & 100 & 77.75 & 1.000 & 75.95 & 1.000 & 80.95 & 1.000 & 71.35 & 1.000 \\
      & 200 & 81.20 & 1.000 & 80.35 & 1.000 & 86.65 & 1.000 & 79.40 & 1.000 \\
    \midrule
    \multirow{6}{*}{PGD-$L_2$}
      & 5   & 17.20 & 0.990 & 15.40 & 0.974 & 11.15 & 0.937 & 11.45 & 0.892 \\
      & 10  & 33.75 & 0.999 & 30.55 & 1.000 & 24.10 & 0.996 & 24.70 & 0.995 \\
      & 20  & 56.25 & 1.000 & 51.95 & 1.000 & 45.80 & 1.000 & 42.10 & 1.000 \\
      & 50  & 75.75 & 1.000 & 74.15 & 1.000 & 74.85 & 1.000 & 66.40 & 1.000 \\
      & 100 & 81.70 & 1.000 & 81.45 & 1.000 & 84.55 & 1.000 & 78.10 & 1.000 \\
      & 200 & 81.50 & 1.000 & 81.90 & 1.000 & 87.40 & 1.000 & 82.60 & 1.000 \\
    \midrule
    \multirow{6}{*}{PGD-$L_\infty$}
      & 5   & 12.95 & 0.996 & 11.50 & 0.988 & 9.30 & 0.866 & 7.95 & 0.904 \\
      & 10  & 27.90 & 1.000 & 24.55 & 1.000 & 20.15 & 0.994 & 19.15 & 0.996 \\
      & 20  & 47.95 & 1.000 & 43.10 & 1.000 & 39.30 & 1.000 & 35.35 & 1.000 \\
      & 50  & 70.60 & 1.000 & 68.00 & 1.000 & 65.50 & 1.000 & 58.65 & 1.000 \\
      & 100 & 77.90 & 1.000 & 74.70 & 1.000 & 78.25 & 1.000 & 71.55 & 1.000 \\
      & 200 & 80.60 & 1.000 & 80.25 & 1.000 & 83.95 & 1.000 & 78.60 & 1.000 \\
    \bottomrule
  \end{tabular}
\end{center}

\begin{center}
  \captionof{table}{Full numerical EER (\%) and RTF results for the purification-step analysis on ECAPA-TDNN under white-box attacks.
  We compare AudioPure and PnP-Diff at different step numbers across MI-FGSM, PGD-$\ell_2$, and PGD-$\ell_\infty$ attacks. RTF is reported in the last column.}
  \label{tab:step-ablation-supp}
  \setlength{\tabcolsep}{3pt}
  \renewcommand{\arraystretch}{1.05}
  \resizebox{\textwidth}{!}{
  \begin{tabular}{l *{6}{c} *{6}{c} *{6}{c} c}
    \toprule
    \multirow{2}{*}{Method} &
    \multicolumn{6}{c}{MI-FGSM} &
    \multicolumn{6}{c}{PGD-$\ell_2$} &
    \multicolumn{6}{c}{PGD-$\ell_\infty$} &
    \multirow{2}{*}{RTF} \\
    \cmidrule(lr){2-7}\cmidrule(lr){8-13}\cmidrule(lr){14-19}
     & 5 & 10 & 20 & 50 & 100 & 200
     & 5 & 10 & 20 & 50 & 100 & 200
     & 5 & 10 & 20 & 50 & 100 & 200
     & \\
    \midrule
    AudioPure-1 & 2.95 & 3.05 & 3.40 & 3.85 & 3.85 & 3.85 & 2.95 & 3.15 & 3.25 & 3.55 & 3.55 & 3.75 & 2.95 & 2.95 & 3.20 & 3.75 & 3.55 & 3.55 & 0.050 \\
    AudioPure-2 & 4.40 & 4.55 & 4.70 & 5.10 & 5.05 & 4.90 & 4.40 & 4.45 & 4.60 & 4.80 & 4.80 & 4.90 & 4.35 & 4.50 & 4.65 & 4.85 & 4.85 & 4.90 & 0.091 \\
    AudioPure-3 & 6.35 & 6.35 & 6.55 & 6.85 & 6.90 & 6.85 & 6.35 & 6.40 & 6.55 & 6.70 & 6.65 & 6.60 & 6.35 & 6.40 & 6.50 & 6.75 & 6.75 & 6.80 & 0.132 \\
    AudioPure-SSNI & 3.30 & 3.45 & 3.85 & 4.05 & 4.10 & 4.05 & 3.20 & 3.50 & 3.70 & 3.70 & 4.05 & 3.70 & 3.20 & 3.40 & 3.55 & 3.75 & 4.00 & 3.90 & 1.085 \\
    \midrule
    PnP-Diff-1 & 2.05 & 2.65 & 3.20 & 4.45 & 3.95 & 3.65 & 2.15 & 2.60 & 3.20 & 3.55 & 3.60 & 3.60 & 2.05 & 2.35 & 3.10 & 3.50 & 3.45 & 3.20 & 0.014 \\
    PnP-Diff-2 & 2.85 & 2.90 & 3.45 & 4.10 & 3.75 & 3.50 & 2.90 & 3.00 & 3.30 & 3.40 & 3.70 & 3.85 & 2.95 & 2.75 & 3.40 & 3.35 & 3.65 & 3.50 & 0.014 \\
    PnP-Diff-3 & 3.25 & 3.50 & 3.85 & 4.30 & 4.05 & 3.90 & 3.25 & 3.55 & 3.70 & 4.00 & 4.00 & 4.10 & 3.25 & 3.60 & 3.80 & 3.95 & 4.00 & 3.85 & 0.014 \\
    PnP-Diff-SSNI & 2.10 & 2.65 & 3.15 & 4.30 & 4.00 & 3.80 & 2.25 & 2.70 & 3.05 & 3.60 & 3.65 & 3.65 & 2.05 & 2.55 & 3.05 & 3.55 & 3.25 & 3.45 & 0.026 \\
    \bottomrule
  \end{tabular}
  }
\end{center}

\begin{center}
  \centering
  \includegraphics[width=0.95\textwidth]{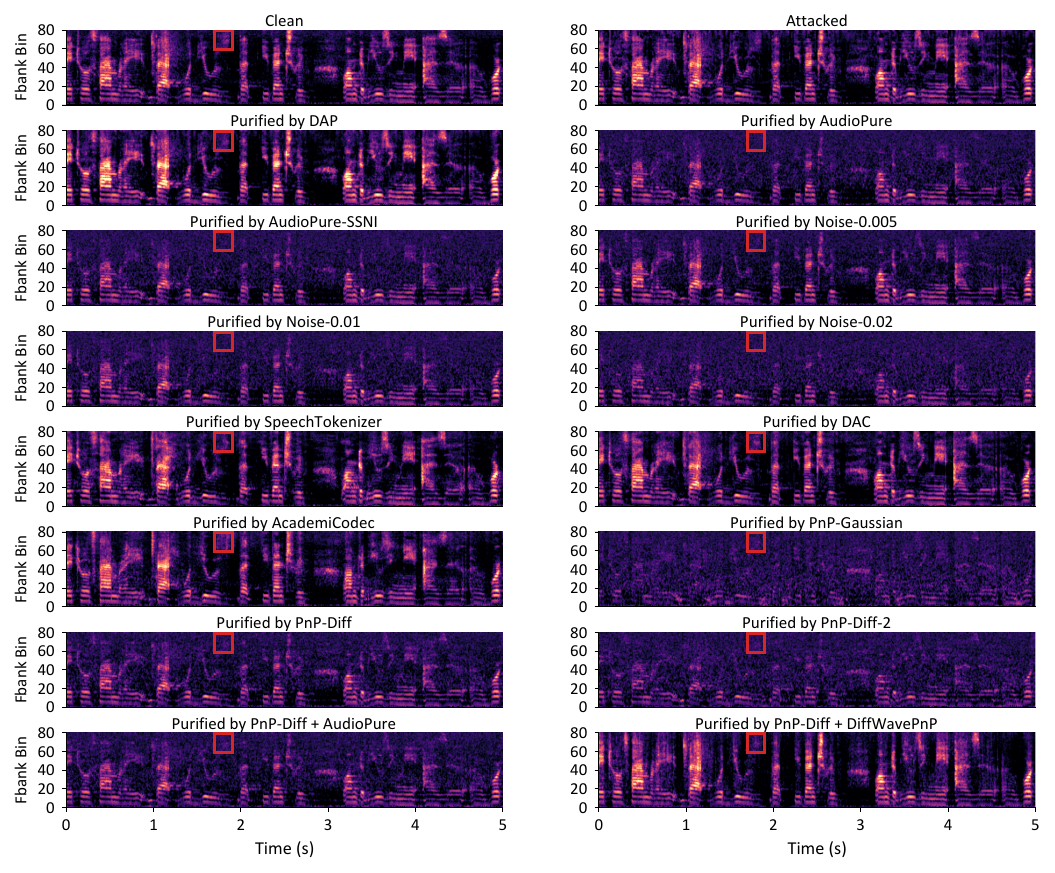}
  \captionof{figure}{Full Fbank comparison under 50-step MI-FGSM for sample \texttt{voxceleb1/id10270/5r0dWxy17C8/00002.wav}. This extended visualization includes all purification methods from the main white-box and black-box comparison table.}
  \label{fig:quality-case-supp}
\end{center}

\vfill

\end{document}